\documentclass{andromedaone}       
\journal{NDM}%Letters in High Energy Physics}
\vol{2020} \jyear{Egypt} \pages{Hurgada} 
\received{xx January 2018}
\published{xx March 2018}
\def\be{\begin{equation}}
\def\ee{\end{equation}}
\def\bea{\begin{eqnarray}}
\def\eea{\end{eqnarray}}

\usepackage{biolinum}
\usepackage{siunitx}
\usepackage{amsmath}
\usepackage{amsmath,amssymb}
\numberwithin{equation}{section}
\begin{document}
\title{Potential of SKA to Detect CDM ALPs with Radio Astronomy}
\author{Ahmed Ayad and Geoff Beck}
\address{School of Physics, University of the Witwatersrand, Private Bag 3, WITS-2050, Johannesburg, South Africa}
%-------------------------------------------------------------------------

\begin{abstract}
Axion-like particles (ALPs) are light pseudo-scalar particles predicted in many theoretically well-motivated extensions to the standard model of particle physics (SM). The search for cold dark matter (CDM) ALPs has gained tremendous ground over the last few years. Essentially ALPs are characterized by their coupling with two photons which allows ALPs to decay into two photons. In this work, we explore the potential of the Square Kilometer Array (SKA) to detect CDM ALPs with radio astronomy in an attempt to detect an observational signature of ALPs conversion onto photons in astrophysical field.
\end{abstract}
\maketitle
\begin{keyword}
dark matter theory, axions, axion-like particles, particle physics-cosmology connection, radio astronomy, the Square Kilometer Array
%\doi{10.2018/LHEP000001}
\end{keyword}

\section{Introduction} \label{sec.1}

It is established from modern cosmology that most of the matter content of the universe is constituted by dark matter (DM) \cite{komatsu2011seven}. However, the exact nature of such a component is not clear yet which provides a central problem for astronomy and cosmology. One theoretically well motivated approach would be to consider light scalar or pseudo-scalar candidates of DM such as axions and axion-like particles (ALPs). Axions \cite{peccei1977cp, weinberg1978new} are pseudo-Nambu-Goldstone bosons that appear after the spontaneous breaking of the Peccei-Quinn symmetry introduced to solve the CP-violation problem of the strong interaction, which represents one of the serious problems in the standard model of particle physics (SM), for a review see reference \cite{peccei2008strong}. Furthermore, ALPs are pseudo-scalar particles predicted in many theoretically well-motivated extensions to the SM \cite{arvanitaki2010string, cicoli2012type, anselm1982second}. Axions and ALPs are characterized by their coupling with two photons. While the coupling parameter for axions is related to the axion mass, there is no direct relation between the coupling parameter and the masses of ALPs. Nevertheless, it is expected that ALPs share similar phenomenology to that of axions. The theory predicts that such axions, or now more generally ALPs, are very light and weakly interacting with the SM particles, see reference \cite{asztalos2006searches}.  For these reasons, it is argued that ALPs are suitable candidates for the DM content of the universe \cite{preskill1983cosmology, abbott1983cosmological, dine1983not, arias2012wispy, ringwald2012exploring, marsh2016axion}.

The coupling between ALPs and photons via a two-photon vertex gives rise to the mixing between them \cite{raffelt1988mixing}. This mixing would lead to the conversion between ALPs and photons $a \rightarrow \gamma$ through the Primakoff effect in the presence of an external electric or magnetic field. It may also allow for the spontaneous decay of the ALPs into two photons $a \rightarrow \gamma + \gamma$. Both of those two mechanisms can be considered as the basis to search for ALPs by explaining a number of astrophysical phenomena or constraining the ALP properties using observations. Using the standard perturbation calculation, the decay time of the very low mass ALPs with very small coupling to photons found to be significantly larger than the age of the universe and therefore can not be counted for producing any observable signature. Since ALPs are identical bosons, they may form a  Bose-Einstein condensate (BEC) with very high occupation numbers. The stimulated decay of ALPs BEC in expanding universe and under the plasma effects can be count for producing observable signature. The main aim of this work is to explore the potential of the Square Kilometer Array (SKA) to detect CDM ALPs with radio astronomy in an attempt to detect an observational signature due to the stimulated decay of ALPs into photons in astrophysical field. 

The structure of this paper is as follows. In sections \ref{sec.2} and \ref{sec.3}, we briefly review the difference between the processes of spontaneous and stimulated decay of ALP into two-photons. In section \ref{sec.4}, we question the capability of the SKA telescopes to detect a signature of ALPs decay. Finally, our conclusion is provided in section \ref{sec.5}.

\section{{Spontaneous decay of ALPs }} \label{sec.2}

Most of the phenomenological implications of ALPs are due to their feeble interactions with the Standard Model (SM) particles that take place through the effective Lagrangian \cite{sikivie1983experimental, raffelt1988mixing, anselm1988experimental}
\begin{equation} \label{eq.1}
\mathrm{\ell}_{a\gamma} = - \frac{1}{4} g_{a\gamma} \mathrm{F}_{\mu \nu} \tilde{\mathrm{F}}^{\mu \nu} a = g_{a\gamma}\, \mathbf{E} \cdot \mathbf{B} \, a \:,
\end{equation}
where $g_{a\gamma}$ is the ALP-photon coupling parameter with dimension of inverse energy, $\mathrm{F}_{\mu \nu}$ and $\tilde{\mathrm{F}}^{\mu \nu}$ represent the electromagnetic field tensor and its dual respectively, and $a$ donates the ALP field. While $\mathbf{E}$ and $\mathbf{B}$ are the electric and magnetic fields respectively. The two-photon ALPs interaction vertex allows for the Primakoff conversion of ALPs into photons in the presence of electric or magnetic field, as well as for the ALPs radiative decay into photons. The majority of the ALPs searches in both the astrophysical environment and in the laboratory are based on these two processes. The spontaneous decay of an ALP with mass $m_a$ proceeds through the chiral anomaly to two photons, each with a frequency $\nu=m_a / 4 \pi$. The perturbative decay time of ALPs can be expressed in term of the ALP mass and ALP-photon coupling as in \cite{kelley2017radio}
\begin{equation} \label{eq.2}
\Gamma_{\text{pert}}^{-1} = \frac{64 \pi}{m_a^3 g_{a\gamma}^2} \:.
\end{equation}
In our previous paper \cite{ayad2019probing} we found astrophysical evidence to constrain the ALPs properties with mass $m_a \sim 10^{-13} \; \text{eV}$ and coupling with photons $g_{a\gamma} \sim 10^{-14} \; \text{GeV}^{-1}$. One can insert these limits to equation (\ref{eq.2}) and evaluate the perturbative decay time for general ALPs which found to be about $\sim 10^{60} \; \text{years}$. This lifetime for ALPs is much larger than the age of the universe and therefore ALPs seem to be super stable on the cosmological scale. Perhaps this is the main reason for neglecting the ALPs decay in the literature. According to this scenario, the spontaneous decay of ALPs can not be responsible for producing any observable signal that can be detectable by the current or near-future radio telescopes. 
  
\section{{Stimulated decay of ALPs }} \label{sec.3}

An important consequence of the context of ALP decay is the fact that ALPs are identical bosons their very low mass indicates that their density and occupation numbers can be very high. Therefore it is suggested that ALPs may form a Bose-Einstein condensate (BEC). This ALPs BEC can then thermalize through their gravitational attraction and self-interactions to spatial localized clumps \cite{chang1998studies, sikivie2009bose}. Of particular importance here is that the system in such high occupancy case is well described by a classical field \cite{guth2015dark}. In such a situation the decay rate receive a Bose enhancement via the photon occupation number $N_k$
\begin{equation} \label{eq.3}
\Gamma= \Gamma_{\text{pert}}(1+2 N_k)  \:.
\end{equation}
Although the spontaneous decay time of ALPs is tremendously larger than the age of the universe, the decay time obtained in \cite{alonso2019wondrous} from the classical equation of motion is just about $\sim 10^{-7} \; \text{s}$, which is dramatically small comparing to the first value. Theoretically, ALPs are expected to be created in the very early universe era, and according to this scenario they have an extremely short lifetime and they supposed to decay so rapidly. Therefore, they must vanish a long time ago and currently, we can not hope for any observable signal to remain due to the stimulated decay of ALPs.

The huge discrepancy between the classical and the standard perturbative calculations can be explained as we ignored the fact that the universe is expanding and full of ionized plasma. In addition to these two reasons, the self-interactions of ALPs condensate may also slow down their decay rate. Considering the very low mass of ALPs and their incredibly weak coupling, the stimulated emission in an empty and non-expanding universe would induce ALPs to decay very rapidly into photons, nullifying most of the interesting parameter space. However, it is strongly argued that ALPs were produced before or during inflation, which implies that they were present during inflation. In this case, the rapid expansion would lead to an extremely homogeneous and coherent initial state. In addition to the expansion of the universe, the plasma effects also are crucial as it modifies the photons propagation and prevents the early decay of ALPs. Inside the plasma, photons have an effective mass that kinematically forbids the decay of lighter particles including ALPs or at least a portion of it. 

Regardless of the role that plasma plays in impeding the early decay of ALPs, we attempt to focus on investigating the plasma effects in modifying the detectability that results from the stimulated emissions from axion clumps. To be more specific, the plasma generates an effective mass for the photons and modifies the dispersion relation of photons. Taking this effect into account would lead to several impacts on the current status of ALPs decay  \cite{alonso2020wondrous}. Therefore we need to resolve the classical equation of motion for photon propagation in an ALP background with the consideration of the plasma effects. Indeed, this is an exciting approach to look for DM ALPs using the current and near-future experiments along with astrophysical observations.

\section{Can SKA telescopes detect signature of ALPs decay?} \label{sec.4}

In the past years, there have been many attempts to detect ALPs DM using radio telescopes, most of which were based on the Primakoff conversion of ALPs into photons \cite{caputo2018looking, caputo2019detecting}. Recently it was shown that the stimulated ALP decays in the astrophysical environments may also generate radio signals comparable with that of the Primakoff conversion \cite{caputo2018looking, sigl2017astrophysical}. As we discussed in the previous section, the stimulated decay of ALPs $a \rightarrow \gamma + \gamma$ with a very high rate is possible in the presence of an electromagnetic wave with a certain frequency. During this process, the electromagnetic wave will be greatly enhanced. The decay of ALPs into photons with taking into account the Bose enhancement would lead to an exponentially increasing growth rate of the occupation number of photons and the subsequent radio wave emission. Since the very early eras of the universe originate, it is filled with very hot and dense plasma that affects the photons propagation. With a complete picture of the environment for the present universe, we need to reconsider the calculations for the radio signals generated from the ALPs decay stimulated by the cosmic microwave background (CMB) photons by taking into account the effective mass for the photon that generated by the plasma effects.

The SKA is considered to be the most sensitive radio telescope ever \cite{colafrancesco2015probing}. This makes it the most aspirant radio telescope to unveil any expected signature of DM decay. Based on the scenario of ALPs decay that we explained above, we explore the potential of the near future radio telescopes, in particular the SKA telescopes, to detect CDM ALPs in an attempt to detect an observational signature of ALPs decay into photons in astrophysical field.
 
\section{CONCLUSIONS} \label{sec.5}

Axion-like particles (ALPs) are a very well-motivated candidate to account for the cold dark matter (CDM) content in the universe. They are light pseudo-scalar particles appears in many extensions to the standard model of particle physics (SM) and are characterized by their coupling to two photons, which give rise to the decay of ALP into two photons. In this work, we discussed the possibility to detect an observable signature produced due to the decay of CDM ALPs. For ALPs with mass $m_a \sim 10^{-13} \; \text{eV}$ and coupling with photons $g_{a\gamma} \sim 10^{-14} \; \text{GeV}^{-1}$, the ALPs are super stable on the cosmological scale and their spontaneous decay can not be responsible for producing any detectable radio signal. Since ALPs are identical bosons, they may form a  Bose-Einstein condensate (BEC) with very high occupation numbers. The stimulated decay of ALPs BEC in expanding universe and under the plasma effects can be count for producing observable radio signals enhanced by several orders of magnitude. The stimulated decay of ALPs BEC arising from the presence of ambient photons results in a large enhancement of the decay rate such as being detectable by the near-future radio observations such as the Square Kilometer Array (SKA). Indeed, this should depend on other parameters like the astrophysical environments and radio telescopes sensitivity. That is worth being a subject of intense future work.

\section*{Acknowledgements}
This work is based on the research supported by the South African Research Chairs Initiative of the Department of Science and Technology and National Research Foundation of South Africa (Grant No 77948). A. Ayad acknowledges support from the Department of Science and Innovation/National Research Foundation (DSI/NRF) Square Kilometre Array (SKA) post-graduate bursary initiative under the same Grant. G. Beck acknowledges support from a National Research Foundation of South Africa Thuthuka (Grant No 117969). %The authors would like also to offer special thanks to Prof. S. Colafrancesco, who, although no longer with us, continues to inspire by his example and dedication to the students he served over the course of his career. 

\bibliographystyle{unsrt}
\bibliography{references}

\end{document}